\documentstyle[pra,preprint,aps]{revtex}

\begin{document}
\draft
\title{Electrodynamics of a two-electron atom with retardation and self-interaction }
\author{Jayme De Luca}
\address{Departamento de F\'{i}sica, Universidade Federal de S\~{a}o Carlos, Rod.\\
Washington Luiz km 235, 13565-905 S\~{a}o Carlos, SP, Brazil}
\date{\today}
\maketitle

\begin{abstract}
We study the linear stability of a circular orbit in a two-electron atom,
with the inclusion of retardation and self-interaction effects. We calculate
all the eigenvalues of the linear stability of the circular orbit, expanded
up to third order in (v/c). We discuss how the linear stability results can
be used to construct a resonant normal form constant of motion, which is
important for the short time stability of the orbits (e. g. emission of
sharp spectral lines). We calculate the magnitudes related to some of those
orbits and compare to the quantum atomic results.
\end{abstract}


\section{Introduction}

We have recently proposed that the retardation of the Coulomb interaction
could stabilize some circular orbits of the helium atom by the existence of
an extra constant provided by a resonant normal form \cite{PRL}. This extra
constant would return the coulombian radial instability to the neighborhood
of the original circular orbit in a short time scale. In this letter we
study in full the linear stability of a circular orbit of a two-electron
atom with the inclusion of retardation and self-interaction. The linear
stability is one of the ingredients in the construction of the resonant
normal form. Another work will be dedicated to the detailed construction of
the resonant normal form, which we briefly discuss here in appendix C in an
intuitive way for completeness.

A circular orbit is one where the two electrons are in the same circular
orbit and in phase opposition, that is, along a diameter\cite{Poirier}. The
center of the two-electron atom is supposed to have a positive charge of
value $Ze,(Z>1/4)$ and we henceforth call it the nucleus. The linearized
dynamics about a circular orbit has one unstable direction, one stable
direction and ten neutrally stable directions\cite{Poirier}. For an
infinitely massive nucleus, a two-electron atom has a six-degree of freedom
Hamiltonian system with only four independent constants of motion\cite
{Gutzwiller}. (Namely, these constants are the energy and the three
components of the total angular momentum). Because there are only four
constants, the Coulombian dynamics in the neighborhood of a generic circular
orbit can in principle be unstable (by lack of constants), and actually is
unstable in this case \cite{Poirier}. Here we show that an extra complex
constant can exist in the neighborhood of a discrete set of circular orbits,
when one includes retardation effects. This provides two extra constants,
and together with the four global constants of the two-electron atom, this
can stabilize the dynamics in the region inside the resonance islands, in a
short time scale. This extra constant is provided by a resonant normal form
(adelphic integral \cite{whittaker}) and requires a resonance to exist \cite
{whittaker,Contopoulos,Gustav,Furta}.

Historically, the understanding of the electrodynamics of a charged particle
interacting with its own electromagnetic field\cite
{Lorentz,Dirac,Rohrlich,Coleman,Jackson} came very late. A classical
solution to the self-interaction of a charged particle of very small radius
is described by the Lorentz-Dirac equation of motion, henceforth called LDE.
The derivation of this equation was first presented in a lecture by Lorentz
in 1906, and first published in 1909\cite{Lorentz,Rohrlich}. Lorentz's
theory had many difficulties\cite{Jackson}, and a major progress came only
in 1938, when Dirac\cite{Dirac} produced a covariant derivation without
mention to the structure of the particle. Dirac was also the first to
recognize and understand the runaway solutions to the LDE\cite{Jackson}.
Dirac's theory still suffered from an arbitrary mass term\cite{Rohrlich}. In
1948, the regularization approach invented to treat the Lamb shift, offered
a satisfactory solution to the divergent mass renormalization\cite{Coleman}.
In this work we consider the electrodynamics of pointlike charges with a
renormalized mass, as described by the Lorentz-Dirac theory\cite
{Rohrlich,Coleman,Jackson}.

Another late development of Maxwell's theory was the work of Page (1918)\cite
{Page} on the expansion of the Li\'{e}nard (1898) and Wiechert (1900)
formula. This formula is complicated because of the retardation constraint,
and one way to convert it into a useful differential equation is to develop
the constraint in a Taylor series. This was done by Page up to the fifth
order in 1918\cite{Page}, who also explored this formula in connection with
self-interaction. We henceforth call the expansion of the
Li\'{e}nard-Wiechert interaction the ``Page series''. Truncated to second
order in $(v/c)$, the Page series describes a Lagrangian interaction. This
Lagragian is the Darwin Lagrangian\cite{Jackson,Darwin,Havas}, which
introduces the first retardation correction to the Coulomb dynamics. The
Darwin Lagrangian is used in quantum mechanics to produce the Breit
operator, which is the generalization of Dirac's equation for two-electron
atoms, correct to second order in $(v/c)$\cite{Breit,Bethe}. The Breit
operator has been used successfully to describe the spin-orbit coupling and
fine structure of helium\cite{Bethe,Trabert}. The third-order term of the
Page series is dissipative, and of course not Lagrangian.

The most studied case of a two-electron atom is the helium atom. The Coulomb
dynamics of helium is not completely integrable, because only three
constants of motion in involution exist for the dynamics\cite
{Gutzwiller,Arnold}. This non-integrability of helium appeared historically
as a hindrance for the early quantization attempts of the Coppenhagen school%
\cite{Van Vleck}. Quite recently, because of the renewed interest in
periodic orbit quantization, much has been learned about the Coulomb
dynamics in helium\cite{Wintgen,Kaneko}. For example, the Coulombian radial
instability is now known to lead to self-ionization, after a long-term
chaotic transient, for most initial conditions\cite{Kaneko}. In section \ref
{discussion} we discuss how the six constants can stabilize the dynamics in
the small region inside the resonance islands, in a short time scale.

The introduction of the first retardation correction to the Coulomb
interaction modifies the dynamics in a quantitative and in a qualitative
way, as far as this work is concerned: The Darwin Lagrangian is rotationally
invariant, which generates an angular-momentum-like constant of motion
according to Noether's theorem\cite{Arnold}. This constant is the angular
momentum of the electrons, with relativistic correction, plus the angular
momentum of the electromagnetic field. There is also an energy-like constant
(because the Darwin Lagrangian is time independent). These are small
perturbations of the four constants of the Coulomb dynamics, and this is the
quantitative change. The qualitative change is the existence of the extra
analytic constant in the neighborhood of some orbits, and we stress that
this is a genuine nonlinear effect, because the frequencies of the
linearized dynamics with retardation depend nonlinearly on the orbit's
frequency . This extra constant appears only after one includes the
retardation effects, which unfold a degeneracy of the Coulomb dynamics. In
this sense, we have included the retardation because it is absolutely
necessary, and not to attain some better precision.

The main motivation of our stability studies was to investigate the possible
existence of an extra analytic constant of the motion in the neighborhood of
some special circular orbits, as described by the Darwin Lagrangian.
According to standard normal form theory\cite
{whittaker,Contopoulos,Gustav,Furta}, the existence of an extra analytic
constant of the motion in the neighborhood of an equilibrium point requires
that the eigenvalues of the linearized dynamics satisfy some resonance
condition. Of course, an extra constant would require a new, special
resonance to be satisfied. We use the results of \cite{Furta} to find where
the extra analytic constant can exist and we show how to construct this
extra constant in appendix C.

In this paper we accomplish two things: First we perform the complete
stability analysis of a generic circular orbit, including retardation and
self-interaction up to third order in $(v/c)$ , which is done here in full
detail for future reference for a generic two-electron atom.. A lot of the
work goes into solving for the linear stability problem, which we call
``variational dynamics'' or sometimes tangent dynamics and we develop a
systematic method to handle the variational equations with inclusion of the
retardation and self-interaction corrections. As an application of the
linear problem that we just solved, we explore some of the very interesting
resulting orbits about which an extra constant is possible. We show that a
degeneracy determines which eigenvalue enters the resonance condition to
produce orbits in the correct atomic magnitude. We discuss mainly the cases
of helium and the $Li^{+}$ ion. In this work we do not prove which
resonances do stabilize the orbit in a linewidth time scale, which is the
subject of another work. Since this requires moving to the rotating frame
and performing long normal form calculations, we chose to devote a separate
work to it\cite{deluca}. To make the present work intuitively complete, we
included in appendix C the main ideas of how to construct the extra
constant, that can possibly secure nonlinear stability. We also discuss (in
Section II) how the resonant terms of the normal form can produce a
deformation of the Coulombian orbit in agreement with a sharp line.

Some cautionary notes are in place before one starts reading this paper:
First, we are studying dynamics in the neighborhood of a circular orbit,
which is also a periodic orbit. One should not think though, that we are
doing periodic orbit quantization\cite{Gutzwiller,Ozorio}. It is known that
in general EBK can not be applied to generic orbits of Coulombian helium
because of the unstable direction\cite{Gutzwiller}. Since we are showing
that the inclusion of retardation stabilizes a discrete set of orbits, it
would be of interest for future research to apply EBK quantization in the
neighborhood of those stable orbits, which would be a nice new application
of normal form theory. Of course one would have to start the quantization
not with the Coulomb Hamiltonian but with the Hamiltonian resulting from the
Darwin Lagrangian, which includes retardation effects. \cite{DarwinHam}

The dynamical system with retardation and self-interaction has nothing to do
with the corresponding quantum system in principle (unless otherwise proved
in the future). It is nevertheless very interesting to compare our stability
results to the quantum results. This comparisom immediately exposes the
relevance of the results to atomic physics and suggests what further results
should be sought by use of nonlinear dynamics. The first striking
coincidence is that our dynamical orbits have energies around the correct
atomic magnitudes and the emited frequencies agree very well with some
frequencies of the spectra of helium and $Li^{+}$. The essential novelty in
the work is that we are discussing for the first time a dynamical system
that can emit a sharp line (to emit a sharp line the orbit must be stable in
a time scale of some $10^{6}$ turns, the type of stability that resonant
normal form can provide). Of course this would suggest a deeper interface
between quantum mechanics and classical electrodynamics with retardation,
and we hope that this interface will become more transparent as we go along
with the research. We do not know at present if it is possible to
approximate all the lines in the spectra of helium and $Li^{+}$ in this way.
In our approximation, we retained only the minimal amount of terms
absolutely necessary to arrive at the results. It comes nevertheless as a
surprise that nonlinear dynamics can produce such a sensible approximation
to some of the sharp frequencies of the spectrum of helium and also to
stabilize circular orbits in the correct atomic energies. Last, in this work
we consider only the interaction of pointlike spinless particles, and we are
not including spin in the dynamical system.

This paper is organized as follows: In part \ref{part II} we discuss the
electromagnetic formulas. In part \ref{coulombic} we consider the Coulomb
stability problem, which is order zero of the Page series formula. In part 
\ref{secondS} we discuss the inclusion of the second-order terms and
consider the resonance condition necessary for an extra constant and compare
some results. In part V we consider the influence of the radiative terms, in
part VI we compare our results to the atomic results and in part \ref
{discussion} we put the discussions.

\section{Electromagnetic formulas}

\label{part II}

In this work we include the self-interaction effects as described by the
relativistic Lorentz-Dirac equation (LDE) with a renormalized mass\cite
{Dirac,Rohrlich,Coleman,Jackson}. The LDE equation for an electron of charge 
$-e$ can be written in the convenient, noncovariant form as 
\begin{equation}
{\frac{{d}}{{dt}}}(\gamma M_{e}\dot{{\bf x}}_{e})=\Gamma +{\bf F}_{ext},
\label{eq:ld}
\end{equation}
where $\dot{{\bf x}}_{e}$ is the electron velocity, $M_{e}$ is the
renormalized electronic mass\cite{Rohrlich} and $\gamma \equiv 1/\sqrt{1-(|%
\dot{{\bf x}}_{e}|/c)^{2}}$. In (\ref{eq:ld}), ${\bf F}_{ext}$ is the
external force acting on the electron and $\Gamma $ is the radiation
reaction force. For circular orbits, the lowest order term of $\Gamma $ in
powers of $(v/c)$ is $\Gamma ={\frac{{2e^{2}}}{{3c^{3}}}}\stackrel{%
\textstyle ...}{\bf x}_{e}$. The next correction to $\Gamma $ is of order $%
(v/c)^{5}$ for a circular orbit\cite{Rohrlich}.

We now introduce, for later use, the expansion of the retardation constraint
of the Li\'{e}nard-Wiechert interaction, (the Page series\cite{Page}). Let $%
{\bf x}$ be the position of a charge $q$, and $\beta $ its velocity vector
divided by $c$. The formula for the electric field caused by this charge $q$
at a point $p$ is 
\begin{eqnarray}
{\bf E}_{p} &=&q{\frac{{\hat{{\bf n}}}}{{r^{2}}}}+q\{{\frac{{(|{\bf \beta }%
|^{2}-3(\hat{{\bf n}}\cdot \beta )^{2})\hat{{\bf n}}}}{{2r^{2}}}}-{\frac{{%
\dot{{\bf \beta }}}}{{2rc}}}-{\frac{{(\hat{{\bf n}}\cdot \dot{\beta})\hat{%
{\bf n}}}}{{2rc}}}\}  \nonumber \\
\mbox{  } &&+{\frac{{2q}}{{3c^{3}}}}\stackrel{\textstyle ...}{\bf x}+\ldots
\label{SchotP-E}
\end{eqnarray}
where $\hat{{\bf n}}$ is the unit vector pointing from the charge to the
point $p$, $r$ is the distance from the charge to $p$ and the ellipsis
represents terms of order higher than 3 in $(v/c)$. Notice that all
functions are evaluated at present time. For a circular orbit, the term of (%
\ref{SchotP-E}) inside braces is of order $(v/c)^{2}$ times the Coulomb term
and the term with the third derivative of position is of order $(v/c)^{3}$
times the Coulomb force. A detailed expansion, correct to fifth order in $%
(v/c)$ is calculated in \cite{Page}. The magnetic field caused by the charge 
$q$ at $p$ has the following series 
\begin{equation}
{\bf B}={\frac{{q}}{{r^{2}}}}[\beta \times \hat{{\bf n}}]+\ldots
\label{SchotP-B}
\end{equation}
This first term is the Biot-Savart term and the next term in the series
would produce a force of fourth order in $(v/c)$ and along the normal, and
it is not important for the present work. In this work we consider only the
above terms of the electromagnetic interaction, and also consider the
relativistic correction of Newton's law for the electronic motion up to
second order in $(v/c)$.

Let us now discuss electrodynamics with retardation and self-interaction in
the special case of the helium atom: We recall that the circular orbit is
defined as one in which the two electrons are in the same circular orbit but 
$180$ degrees out of phase\cite{Poirier}. Along such orbit, the total force
acting on electron 2 can be calculated using (\ref{SchotP-E}) and the
self-interaction of electron 2 to be 
\begin{equation}
{\bf F}={\frac{{2e^{2}}}{{3c^{3}}}}(\stackrel{\textstyle ...}{\bf x}_{1}+%
\stackrel{\textstyle ...}{\bf x}_{2})-7{\frac{{e^{2}{\bf \hat{n}}}}{{4R^{2}}}%
}(1-{\frac{{3}}{{7}}}|\beta |^{2}),  \label{eq:LDDL}
\end{equation}
where $R$ is the radius of the circular orbit. In (\ref{eq:LDDL}) we have
also added the Coulomb attraction of the $\alpha $ particle and used the
approximation that the $\alpha $ particle is infinitely massive and resting
at the origin. If the two electrons are in the same circular orbit but in
phase opposition, the force along the velocity cancels out (first term on
the right of equation \ref{eq:LDDL}), demonstrating that the circular orbit
is a possible periodic solution of the electromagnetic equations up to third
order.

Notice the appearance of the dipole term in (\ref{eq:LDDL}), 
\begin{equation}
\ddot{{\bf D}}=-e(\ddot{{\bf x}}_{1}+\ddot{{\bf x}}_{2}).  \label{eq:dipole}
\end{equation}
The total far-field caused by the three particles depends linearly on the
quantity $\ddot{{\bf D}}$ defined by equation (\ref{eq:dipole}), up to
quadrupole terms. If this quantity is zero, the orbit is not radiating in
dipole. The fifth-order terms of the Page series force and the fifth-order
relativistic correction to the Lorentz-Dirac self-interaction introduce
quadrupole effects. These effects would be important only in a much longer
time scale, of order $T/(v/c)^{5}$. In section VI we show that the circular
orbits decay in a time of the order of $T/(v/c)^{3}$. Therefore, the effect
of quadrupole terms is small during the orbits lifetime. We start the next
paragraph by discussing this importance of quadrupole terms in detail.

Throughout this paper, it is very important to keep in mind the orders of
magnitude relevant to atomic physics. For example, a typical value for $%
(v/c) $ is $(v/c)\sim 10^{-2}$. Typical values for the width of the spectral
lines is of the order of $(v/c)^{3}/T$, which is the inverse of a time to
perform about $10^{6}$ turns along a typical orbit\cite{Seidl}. In the
classical model of the isolated hydrogen atom, because of dipole radiation
losses, the energy loss during this linewidth-time produces dramatic changes
in the frequency of rotation and therefore a band of dipole radiation is
emitted, not a sharp line\cite{Seidl,braz}. We explore this in another
publication\cite{braz}, and we mention here this classical argument of Bohr%
\cite{Bohr} just to stress that it does not apply for the circular orbits of
a two electron atom. Along circular orbits, because there is no dipole
radiation, a sharp line can be emitted for motion along the stable manifold%
\cite{Guckenheimer}, provided the orbit is otherwise stable in an
intermediate time scale. The quadrupole power radiated is of size ($%
E_{o}/T)(v/c)^{6},$ where $E_{o}$ is the coulombian energy of the orbit and
T its period. This power times the linewidth time $T/(v/c)^{3}$ results in
an energy loss of $E_{o}(v/c)^{3},$ which is consistent with a sharp
variation of the emitted frequency. The circular orbits should decay to the
ground state or ionize in a time corresponding to the inverse of the
linewidth, and for us here it is only important the fact that they can
radiate a sharp frequency in the process.

A word of caution should be said about the fact that a term in the Page
series of order higher than two represents a singular perturbation.
Therefore, when one such term is included in the equations of motion, there
will be solutions to the dynamics that are not a perturbation of a
mechanical coulombian orbit. We call these solutions ``non-mechanical'', as
opposed to ''quasi-mechanical'' regular perturbations. In this work we
investigate only quasi-mechanical regular perturbations of circular orbits.

If one wants the most generic ``stationary-state'', where the helium atom
does not radiate in dipole, then one must have $\ddot{{\bf D}}=-e({\bf \ddot{%
x}}_{1}+{\bf \ddot{x}}_{2})=0$ for all times. If we integrate the condition $%
\ddot{{\bf x}}_{1}+\ddot{{\bf x}}_{2}=0$ twice in time, we get 
\begin{equation}
{\bf x}_{1}+{\bf x}_{2}=a+bt.  \label{sumi}
\end{equation}
The constant $b$ must be zero if the electrons are bound to the center of
force at the origin. Inspection shows that $a$ must also be zero, which is
necessary for the existence of a solution to the equations of motion with
the Coulomb force terms. Along orbits that satisfy (\ref{sumi}) with $a$ $%
=b=0$ , the interaction between the electrons just renormalizes the charge
at the center of force. The most general ``quasi-mechanical'' stationary
orbit possible is then an elliptical orbit, with circular orbits as a
particular case. In this work we do not consider elliptical orbits, but they
could be studied in a way analogous to the circular orbits. The other kind
of possible stationary orbits are symmetric collinear motions of the two
electrons. These are orbits of zero angular momentum and they are also
singular orbits (Coulombian solutions with zero angular momentum and zero
dipole would fall onto the nucleus). We discuss them briefly in section \ref
{discussion}.

If one is not interested in non-mechanical orbits, it is convenient to
truncate the Page series to third order and assume that the truncated system
describes all the essential electrodynamics and that the next terms only
introduce a small stochasticity. For non-mechanical orbits, such as zero
angular momentum orbits, the Page series is not convergent and it might be
necessary to keep all orders, like in Eliezer's theorem\cite
{braz,Eliezer,Parrott}. In the case the series is at least asymptotic, the
successive orders become important in successively longer time scales. For
example, about a periodic orbit of period $T$, the second order terms
produce deviations from the Coulomb dynamics in a time of order $T/(v/c)^{2}$%
. The third order terms take a time $T/(v/c)^{3}$ to influence the dynamics,
and so on. In this work we consider the well defined dynamical system
obtained by truncating the Page series interaction to third order, and we
include the self-interaction to third order in $(v/c)$ as well.

\section{Coulombian stability of circular orbits}

\label{coulombic}

In this part we consider the non-relativistic Coulomb dynamics of a
two-electron atom in the neighborhood of a circular orbit. The plain Coulomb
stability problem has already been considered by many authors\cite
{Poirier,Nicholson} using other approaches, and we do it again to introduce
our perturbation scheme. The Coulomb interaction is order zero of the Page
series interaction (\ref{SchotP-E}), and the scheme we develop allows for
the inclusion of higher order terms of the interaction in an easy way. An
alternative equivalent way to perform this calculation is to transform to a
coordinate system rotating with the frequency of the circular orbit. In this
system the circular orbit is a fixed point of an autonomous vector field,
the Jacobian matrix is independent of time, and the parametric problem is
replaced by a linear eigenvalue problem. The disadvantage is that one has to
transform all the terms of the Page series to the rotating coordinates. We
get back to rotating coordinates later on.

Newton's equations of motion for the Coulombian two-electron atom are 
\begin{eqnarray}
M_{\alpha }\ddot{{\bf x}}_{\alpha } &=&{\frac{{Ze^{2}}}{{R_{1\alpha }^{3}}}}(%
{\bf x}_{1}-{\bf x}_{\alpha })+{\frac{{Ze^{2}}}{{R_{2\alpha }^{3}}}}({\bf x}%
_{2}-{\bf x}_{\alpha }),  \nonumber \\
M_{e}\ddot{{\bf x}}_{1} &=&-{\frac{{Ze^{2}}}{{R_{1\alpha }^{3}}}}({\bf x}%
_{1}-{\bf x}_{\alpha })-{\frac{{e^{2}}}{{R_{12}^{3}}}}({\bf x}_{2}-{\bf x}%
_{1}),  \nonumber \\
M_{e}\ddot{{\bf x}}_{2} &=&-{\frac{{Ze^{2}}}{{R_{2\alpha }^{3}}}}({\bf x}%
_{2}-{\bf x}_{\alpha })-{\frac{{e^{2}}}{{R_{12}^{3}}}}({\bf x}_{1}-{\bf x}%
_{2}).  \label{newtons}
\end{eqnarray}
where ${\bf x}_{\alpha }$, ${\bf x}_{1}$ and ${\bf x}_{2}$ are the position
vectors respectively of the nucleus, electron 1 and electron 2, $%
R_{12}\equiv |{\bf x}_{1}-{\bf x}_{2}|$, $R_{1\alpha }\equiv |{\bf x}_{1}-%
{\bf x}_{\alpha }|$ and $R_{2\alpha }\equiv |{\bf x}_{2}-{\bf x}_{\alpha }|$%
. In the special case of helium, the nucleus is an $\alpha $ particle, but
we will use the index $\alpha $ to label the coordinate of the nucleus in
the generic case as well . The nucleus has an arbitrary charge of $Ze$, and
the electrons have charge $-e$. The circular orbit periodic solution of (\ref
{newtons}) is 
\begin{eqnarray}
x_{\alpha }=0, &y_{\alpha }=0,&z_{\alpha }=0,  \nonumber \\
x_{1}=R\cos (\omega t), &y_{1}=R\sin (\omega t)&,z_{1}=0,  \nonumber \\
x_{2}=-R\cos (\omega t), &y_{2}=-R\sin (\omega t),&z_{2}=0.  \label{rotsol}
\end{eqnarray}
According to (\ref{newtons}), the frequency of the orbit is related to $R$
by 
\begin{equation}
M_{e}\omega _{o}^{2}=(Z-\frac{1}{4}){\frac{{e^{2}}}{{R^{3}}}}.
\label{Coulombdisper}
\end{equation}
In this section, $\omega $ and $\omega _{o}$ are the same, but we will see
later on that because of higher order corrections, $\omega _{o}$ is only the
first term of $\omega $ in powers of $(v/c)^{2}$. To simplify the notation
we define the quantity 
\[
\phi \equiv \left( \frac{1}{8Z-2}\right) , 
\]
which specifies a generic two-electron atom.

Linearizing (\ref{newtons}) about the circular orbit (\ref{rotsol}), we
obtain a parametric linear differential equation with coefficients periodic
in time and period $T=\pi /\omega $, as we show in appendix \ref{appA}. If
the Floquet exponents are all nondegenerate, one can find a complete set of
solutions of the form\cite{Jordan} 
\begin{eqnarray}
\delta {\bf x}_{\alpha } &=&\exp (2i\omega \mu t)\sum_{n}{\bf x}_{n}^{\alpha
}\exp (2in\omega t),  \nonumber \\
\delta {\bf x}_{1} &=&\exp (2i\omega \mu t)\sum_{n}{\bf x}_{n}^{1}\exp
(2in\omega t),\   \nonumber \\
\delta {\bf x}_{2} &=&\exp (2i\omega \mu t)\sum_{n}{\bf x}_{n}^{2}\exp
(2in\omega t),  \label{Floquet}
\end{eqnarray}
where the Floquet exponent $\mu $ is a complex number defined in the first
Brillouin zone, $-1/2<Re(\mu )<1/2$. From now on, an upper index should not
be confused with an exponent, and takes values $\alpha $, $1$, and $2$ to
label the nucleus, electron 1, and electron 2, respectively. Notice that for
the Floquet components we use an upper index, but to label coordinates as
functions of time, as in (\ref{newtons}), we use a lower index (to
distinguish it from the Floquet components). To bring the variational
equations to normal form, we define the coordinates 
\[
\xi _{n}^{\kappa }\equiv {\frac{{1}}{{\sqrt{2}}}}(x_{n}^{\kappa
}-iy_{n}^{\kappa }),\mbox{  }\chi _{n}^{\kappa }\equiv {\frac{{1}}{{i\sqrt{2}%
}}}(x_{n}^{\kappa }+iy_{n}^{\kappa }), 
\]
where again the upper index is not to be confused with an exponent, and
takes the values $\alpha $, $1$, and $2$. Next we calculate Hill's secular
determinant\cite{Jordan}, which reduces to the evaluation of a $6\times 6$
determinant in this case of circular orbits. As a simplification, let us
define $\bar{n}\equiv n+\mu $ to be the running variable in the summations
of (\ref{Floquet}). Notice that $\bar{n}$ defines a frequency of linear
oscillation in units of $2\omega $, as of (\ref{Floquet}). Last, to
introduce the physical intuition in the problem and explore the symmetries,
it is convenient to define the coordinates ${\bf x}_{r}$ and ${\bf x}_{d}$
as 
\begin{eqnarray}
{\bf x}_{r} &\equiv &{\bf x}_{1}+{\bf x}_{2}-2{\bf x}_{\alpha },  \nonumber
\\
{\bf x}_{d} &\equiv &{\bf x}_{1}-{\bf x}_{2},  \label{defradrel}
\end{eqnarray}
respectively, which introduces two new letters, $r$ and $d$ to appear as
labels. In helium, $(Z=2)$, the coordinate $\delta {\bf x}_{r}$ is the
dipole moment, and a nonzero amplitude for this variation will always mean
that the normal mode radiates in dipole. The dynamics of the $\delta {\bf x}%
_{d}$ variation is decoupled from the $\delta {\bf x}_{\alpha }$ and $\delta 
{\bf x}_{r}$ variations about circular orbits. Before we write the
equations, we need still another definition 
\begin{eqnarray}
U_{n}^{\kappa } &\equiv &{\frac{{1}}{{\sqrt{2}}}}\xi _{n-1}^{\kappa }+{\frac{%
{i}}{{\sqrt{2}}}}\chi _{n+1}^{\kappa }  \nonumber \\
V_{n}^{\kappa } &\equiv &{\frac{{-i}}{{\sqrt{2}}}}\xi _{n-1}^{\kappa }-{%
\frac{{1}}{{\sqrt{2}}}}\chi _{n+1}^{\kappa }.  \label{defUL}
\end{eqnarray}

It is convenient to think of the quantities $U_{n}^{\kappa }$ and $%
V_{n}^{\kappa }$ as the x and y components of a two-dimensional vector ${\bf %
K}_{n}^{\kappa }$. At this point one should take a look at Appendix \ref
{appA}, where we write the variation of the useful functional forms in terms
of the vector Floquet components. Let us start by considering variations
along the plane of the orbit. Using the results of Appendix \ref{appA}, the
planar variational equations of equation (\ref{newtons}) can be written most
simply in terms of the Floquet components of (\ref{Floquet}) as 
\begin{eqnarray}
-4\bar{n}^{2}{\bf x}_{n}^{d}-{\frac{{1}}{{2}}}({\bf x}_{n}^{d}+3{\bf K}%
_{n}^{d}) &=&0,  \nonumber \\
-4\bar{n}^{2}{\bf x}_{n}^{\alpha }+{4Z\phi }({\bf x}_{n}^{r}+3{\bf K}%
_{n}^{r})/\varrho &=&0,  \nonumber \\
-4\bar{n}^{2}{\bf x}_{n}^{r}-{4Z\phi }(1+{\frac{{2}}{{\varrho }}})({\bf x}%
_{n}^{r}+3{\bf K}_{n}^{r}) &=&0,  \label{Wannierinst}
\end{eqnarray}
where $\varrho $ is defined as the ratio of the mass of the nucleus to the
electronic mass. For example, in the case of helium $\varrho \approx 7344$ .
For the Coulomb dynamics only, also the last equation of (\ref{Wannierinst})
depends only on the radiation coordinate. It is good to have a scheme in
mind to keep track of what we have done so far: equation (\ref{Wannierinst})
is the variation of (\ref{newtons}) divided by $M_{e}\omega _{o}^{2}$, which
we write schematically as 
\[
\frac{M\delta {\bf A}^{(0)}-\delta {\bf F}^{(0)}}{M_{e}\omega _{o}^{2}}=0. 
\]
In part IV we consider the variation of the relativistic corrections to the
acceleration, which we call $\delta M{\bf A}^{(2)}$, and the Page series
second order force, which we call $\delta {\bf F}^{(2)}$. The equations of
motion will then be written schematically as 
\[
\frac{M\delta {\bf A}^{(0)}-\delta {\bf F}^{(0)}}{M_{e}\omega _{o}^{2}}+%
\frac{\delta M{\bf A}^{(2)}-\delta {\bf F}^{(2)}}{M_{e}\omega _{o}^{2}}%
+\ldots =0. 
\]
In this way, we keep adding higher order matrices and calculating Hill's
secular determinant up to an order. Since we will be adding the matrices, it
is necessary to keep some convention about the order in which the equations
appear. The convention is that we always repeat what we did in (\ref
{Wannierinst}), that is, first the equation of motion for $\delta {\bf x}%
_{d} $ divided by $M_{e}\omega _{o}^{2}$, then the equation of motion for
the $\delta {\bf x}_{\alpha }$ divided by $M_{\alpha }\omega _{o}^{2}$ and
then the equation for $\delta {\bf x}^{r}$ divided by $M_{e}\omega _{o}^{2}$.

We recall that $\bar{n}=n+\mu $, and there is one equation for every value
of $n$. In principle this would lead to an infinite Hill's secular
determinant, which is the case for an elliptic periodic orbit. For the case
of circular orbits, when we go to the $\xi $ and $\chi $ variables, we find
that the $\xi _{n}$ variables couple only to $\chi _{n+1}$ and vice versa.
This is obtained by taking linear combinations of the $x$ and $y$ components
of the vectorial equations (\ref{Wannierinst}), a procedure which we call
``unvectorizing''. We developed this convenient mnemonic method of sketching
the calculation to avoid making mistakes in an otherwise lenghty algebra
(even though we did it with the symbolic manipulator Maple). The resulting
matrix equations are 
\begin{equation}
\left[ 
\begin{array}{cc}
-({\frac{{1}}{{2}}}+4\bar{n}^{2}) & {\frac{{-3i}}{{2}}} \\ 
{\frac{{3i}}{{2}}} & -({\frac{{1}}{{2}}}+4\bar{n}_{+}^{2})
\end{array}
\right] \left[ 
\begin{array}{c}
\xi _{n}^{d} \\ 
\chi _{n+1}^{d}
\end{array}
\right] =0,  \label{coul0d}
\end{equation}
and 
\begin{equation}
4\left[ 
\begin{array}{cccc}
-\bar{n}^{2} & 0 & {Z\phi y} & {3iZ\phi } \\ 
0 & -\bar{n}_{+}^{2} & -{3iZ\phi } & {Z\phi y} \\ 
0 & 0 & -(\bar{n}^{2}+\Delta ) & -{3i\Delta } \\ 
0 & 0 & {3i\Delta } & -(\bar{n}_{+}^{2}+\Delta )
\end{array}
\right] \left[ 
\begin{array}{c}
\xi _{n}^{\alpha } \\ 
\chi _{n+1}^{\alpha } \\ 
\xi _{n}^{r} \\ 
\chi _{n+1}^{r}
\end{array}
\right] =0,  \label{coul04}
\end{equation}
where $\bar{n}_{+}\equiv \bar{n}+1$, $y\equiv (1/\varrho )$ and $\Delta
\equiv {Z\phi (1+2y)}$. Finally, let us solve (\ref{coul0d}) and (\ref
{coul04}) for the roots $\bar{n}$. For (\ref{coul0d}) the roots are 
\[
\bar{n}=-1,0,-1/2,-1/2. 
\]
Notice that the secular problems of (\ref{coul0d}) or (\ref{coul04}) involve
only $\bar{n}$ and $\bar{n}_{+}\equiv \bar{n}+1$. Once $\bar{n}$ is
calculated from the secular condition, we can always recover $n$ and $\mu $
because any complex number can be decomposed in a unique way as $\bar{n}%
=n+\mu $ with $n$ integer and $|Re(\mu )|<(1/2)$. After we find $n$, this
defines that the only two nonzero components are the $n$th and $(n+1)$th in (%
\ref{Floquet}) for the normal mode oscillation. An eigenvector calculation
should then follow to determine the ratio of these two Floquet components.

As regards the eight roots of equation (\ref{coul04}), four roots are $%
0,0,-1,-1$. From the general theory of linear ODE's, at a double root like $%
\bar{n}=0$, the general solution is a ``quasi-polynomial'' linear function
of time \cite{Jordan}. The doubly degenerate root $\bar{n}=0$ is then
responsible for the homogeneous translation solution with a constant
velocity of the coulombian helium. Evaluating the determinant of (\ref
{coul04}) and equating it to zero, one finds that the other roots are given
by 
\[
\bar{n}=-{\frac{{1}}{{2}}}+\epsilon , 
\]
where $\epsilon $ is a solution of 
\begin{equation}
\epsilon ^{4}+\left( 2\Delta -\frac{1}{2}\right) \epsilon ^{2}+(\frac{1}{4}%
-2\Delta )(\frac{1}{4}+4\Delta )=0,  \label{Wanieq}
\end{equation}
which is a quartic equation with solutions 
\[
\epsilon ^{2}=\frac{1}{4}-\Delta \pm \sqrt{\Delta (9\Delta -1)}. 
\]
Inspection of the above formula shows that there is always a pair of roots
with negative $\epsilon ^{2}$ and a real pair of roots. The pair with
negative $\epsilon ^{2}$ describes an instability, which was first found by
Nicholson\cite{Nicholson} and in this work we refer to it as the Coulombian
radial instability\cite{Poirier,Kaneko}. This is an exponential growth in a
time scale of the order of one cycle. The corresponding roots are given by 
\begin{equation}
\bar{n}=-{\frac{{1}}{{2}}}\pm i\sqrt{\Delta -\frac{1}{4}+\sqrt{\Delta \left(
9\Delta -1\right) }},  \label{negWan}
\end{equation}
where $i\equiv \sqrt{-1}$. One also finds that $\delta {\bf x}_{r}$ is not
zero for the radially unstable mode, which implies that perturbations along
this unstable mode radiate in dipole. We discuss some consequences of this
in part V. The other two roots of (\ref{Wanieq}), with positive $\epsilon
^{2}$ are approximated by 
\begin{equation}
\bar{n}=-{\frac{{1}}{{2}}}\pm \sqrt{\frac{1}{4}-\Delta +\sqrt{\Delta
(9\Delta -1)}}.  \label{posiWan}
\end{equation}
Notice that the quantity $\zeta $ that appears in the paper of Poirier \cite
{Poirier} is equal to 8 times our $\Delta $ $(\Delta \equiv Z(1+2y)/(8Z-2))$
and that we are including the dynamics of the nucleus as well. To recover
the results of Poirier one should put $y=0$ in our definition of $\Delta $.

Last, we consider the oscillations perpendicular to the plane of the orbit,
which we call the $z$ direction. In linear order in the oscillation
amplitude, this oscillation is decoupled from the oscillations along the
plane. Starting from equation (\ref{newtons}), and linearizing about (\ref
{rotsol}) we obtain 
\begin{eqnarray}
{\frac{{\delta \ddot{z}_{d}}}{{\omega _{o}^{2}}}}+\delta z_{d} &=&0, 
\nonumber \\
{\frac{{\delta \ddot{z}_{\alpha }}}{{\omega _{o}^{2}}}}-{8Z\phi y}\delta
z^{r} &=&0,  \nonumber \\
{\frac{{\delta \ddot{z}_{r}}}{{\omega _{o}^{2}}}}+8\Delta \delta z_{r} &=&0.
\label{lizet}
\end{eqnarray}
The solutions to this linear problem are of type 
\begin{equation}
\delta z_{\kappa }=c_{\kappa }\exp (2i\omega \lambda ).  \label{deflamb}
\end{equation}
The first equation is a simple separate linear equation and the solution to
it with $c_{d}\neq 0$ requires $\lambda =\pm 1/2$. The next two equations of
(\ref{lizet}) become 
\begin{equation}
\left[ 
\begin{array}{cc}
-4\lambda ^{2} & -{8Z\phi y} \\ 
0 & -4\lambda ^{2}+8\Delta
\end{array}
\right] \left[ 
\begin{array}{c}
c_{\alpha } \\ 
c_{r}
\end{array}
\right] =0,  \label{z0}
\end{equation}
and a nontrivial solution requires $\lambda =0,0$ and 
\begin{equation}
\lambda =\pm \sqrt{2\Delta }.  \label{freq27}
\end{equation}

The linear $z$ oscillation is decoupled from the planar oscillation , but of
course it couples at higher orders in the oscillation amplitudes.

As we mentioned in the beginning of this section, in a frame rotating with
the frequency of the circular orbit, the circular orbit itself is a fixed
point of an autonomous vector field. This vector field describes the Coulomb
dynamics in the rotating frame. One finds that the inclusion of the second
order and radiative terms still yields an autonomous vector field for the
dynamics in the rotating frame. This is the reason why we were able to
simplify the parametric equation down to a $6\times 6$ linear system. In
this rotating system one also has to diagonalize a $6\times 6$ matrix (two
plane coordinates for each particle), and the exact same problem along the $%
z $ direction.

\section{Inclusion of Second-Order Terms}

\label{secondS} In this Section we consider the inclusion of the
second-order terms of the Page series and the second-order relativistic
corrections to the electronic dynamics. It is known that the
Li\'{e}nard-Wiechert interaction truncated to second order in (v/c) is
described by the Darwin Lagrangian\cite{Jackson} 
\begin{eqnarray}
&&L_{Darwin}=\sum_{i}({\frac{{1}}{{2}}}m_{i}|\dot{{\bf x}}_{i}|^{2}+{\frac{{1%
}}{{8c^{2}}}}m_{i}|\dot{{\bf x}}_{i}|^{4})  \nonumber \\
&&-{\frac{{1}}{{2}}}\sum_{ij}{\frac{{q_{i}q_{j}}}{{r_{ij}}}}(1-{\frac{{1}}{{%
2c^{2}}}}[\dot{{\bf x}}_{i}\cdot \dot{{\bf x}}_{j}+(\dot{{\bf x}}_{i}\cdot 
\hat{{\bf n}}_{ij})(\dot{{\bf x}}_{j}\cdot \hat{{\bf n}}_{ij})]),
\label{Darwin}
\end{eqnarray}
where the indices take the values 1,2, and $\alpha $, $r_{ij}\equiv |{\bf x}%
_{i}-{\bf x}_{j}|$ and $\hat{{\bf n}}_{ij}$ is the unit vector in the
direction of ${\bf x}_{i}-{\bf x}_{j}$. This Lagrangian depends only on the
scalar product of the velocities and the distance between the particles,
therefore it is invariant under a global rotation of all the particles
coordinates. By Noether's theorem\cite{Arnold}, this generates an
angular-momentum-like constant of the motion associated with the symmetry,
which is equal to the angular momentum plus a small functional correction of
order $(v/c)^{2}$. Because the Lagrangian is time-independent, there is also
an energy constant of the motion. This makes four independent constants of
the motion. According to a recent result on non-integrability of generic
systems of ODE's, for an extra analytic constant to appear, some extra
resonance condition must be satisfied by the linear frequencies\cite{Furta}.

Next we determine the second-order correction of all the Coulomb
eigenvalues. Let us start by correcting the frequency of the orbit as given
by (\ref{Coulombdisper}). A given circular orbit is characterized by the
velocity $|\beta |,$ and from this one can calculate all the other
quantities of the orbit: angular momentum, frequency and radius. The Page
series gives a correction in powers of $|\beta |,$ and the first correction
is of order $|\beta |^{2}$, as it should be for any relativistically
invariant dynamics\cite{Havas}. Adding up the second-order magnetic (\ref
{SchotP-B}) and second-order electric (\ref{SchotP-E}) forces acting on the
electron along the circular orbit we find the normal force 
\[
{\bf F}=-{\frac{{e^{2}{\bf n}}}{8\phi R{^{2}}}}-{\frac{{|\beta _{e}|^{2}e^{2}%
{\bf n}}}{{8R^{2}}}}. 
\]
Together with the second order relativistic mass correction, this determines
the second order correction to the frequency of (\ref{Coulombdisper}) to be 
\[
\omega ^{2}=\omega _{o}^{2}\left[ 1+(\phi -\frac{1}{2})|\beta |^{2}\right] . 
\]
A simple way to include this second-order frequency correction is to
multiply $\bar{n}$ and $\bar{n}_{+}$ by $\left[ 1+\left( \phi -\frac{1}{2}%
\right) |\beta |^{2}\right] $ on the zero-order matrices of (\ref{coul0d})
and (\ref{coul04}) before adding the other second-order matrices. At this
point one should look up appendix B, where we evaluate the variation of the
second-order terms of the Page series, as well as second-order relativistic
corrections.

Adding up the second-order electric field, second-order magnetic force, the
second-order relatistic correction for the electronic masses and the
correction to the frequency, we obtain a perturbation to the matrices (\ref
{coul0d}) and (\ref{coul04}). We write the equation for the Floquet
components, which separates in two parts just like (\ref{Wannierinst})
splits into (\ref{coul0d}) and (\ref{coul04}). The resulting second-order
correction to be added to (\ref{coul0d}) is 
\begin{equation}
|\beta |^{2}\left[ 
\begin{array}{cc}
P_{2}(\bar{n}) & iQ_{2}(\bar{n}) \\ 
-iQ_{2}(-\bar{n}_{+}) & P_{2}(-\bar{n}_{+})
\end{array}
\right] \left[ 
\begin{array}{c}
\xi _{n}^{d} \\ 
\chi _{n+1}^{d}
\end{array}
\right] ,  \label{secd}
\end{equation}
where $P_{2}(\bar{n})\equiv \left[ \left( 8\phi -2\right) \bar{n}^{2}+8\phi 
\bar{n}+\frac{3\phi }{2}\right] $, $Q_{2}(\bar{n})\equiv \left[ \left( 4\phi
+2\right) (\bar{n}^{2}-\bar{n})+\frac{\phi }{2}\right] $ and $\bar{n}%
_{+}\equiv (1+\bar{n})$. Let us first consider the correction in powers of $%
|\beta |$ of the roots of (\ref{coul0d}). This is done by adding up matrices
(\ref{coul0d}) and (\ref{secd}), and taking the determinant of the result.
The result, up to second order in $|\beta |$ is 
\begin{equation}
\begin{array}{l}
16\bar{n}(1+\bar{n})(1+2\bar{n})^{2}+|\beta |^{2}\{\left( 7+6\phi \right)
+8\left( 1-5\phi \right) \bar{n} \\ 
+8\left( 3-13\phi \right) \bar{n}^{2}+16\left( 1-4\phi \right) \left( 2\bar{n%
}^{3}+\bar{n}^{4}\right) \}+...
\end{array}
\label{polyd}
\end{equation}
In the neighborhood of the simple root $\bar{n}=0$ of (\ref{coul0d}), (\ref
{polyd}) takes the form 
\[
16\bar{n}+\left( 7+6\phi \right) |\beta |^{2}=0, 
\]
which yields $\bar{n}=-\frac{\left( 7+6\phi \right) |\beta |^{2}}{16}$.
Analogously, in the neighborhood of $\bar{n}=-1$, (\ref{polyd}) takes the
form 
\[
-16(\bar{n}+1)+\left( 7+6\phi \right) |\beta |^{2}=0, 
\]
which yields $\bar{n}=-1+\frac{\left( 7+6\phi \right) |\beta |^{2}}{16}$.
Last, and crucial for this work, is the bifurcation of the $\bar{n}=-1/2$
double root. About $n=-1/2$, (\ref{polyd}) becomes 
\[
-4(1+2\bar{n})^{2}+6\left( 1+2\phi \right) |\beta |^{2}=0, 
\]
with roots 
\begin{equation}
\bar{n}=-{\frac{{1}}{{2}}}\pm |\beta |\sqrt{\frac{3(1+2\phi )}{2}}.
\label{bifline}
\end{equation}
All the above calculations can be done by hand, and we checked it using the
program Maple, version 4.0. Notice that the correction of the degenerate
root in (\ref{bifline}) comes with a ${\em linear}$ power of $|\beta |$,
differently from the simple roots, that are corrected only at order $|\beta
|^{2}$(exactly because of this degeneracy). This root undergoes the fastest
change of all for small $(v/c)$, and one should expect it to be the first to
accommodate a ``new'' resonance condition. In the construction of a resonant
normal form constant, this frequency allows the resonance condition to be
satisfied with the lowest possible value of $|\beta |$. (In the spirit of
Section VI, one easily finds by inspection that resonances among frequencies
corrected only at second order lead to relativistic orbits, not very
interesting for atomic physics).

Next we calculate the second-order corrections of (\ref{coul04}): Adding up
all the second-order corrections to (\ref{coul04}), second-order electric,
second-order magnetic, second-order relivistic and correction of the
frequency we find the following matrix to be added to (\ref{coul04})

\[
|\beta |^{2}\left[ 
\begin{array}{cccc}
\left( 2-4\phi \right) \bar{n}^{2} & 0 & 0 & 0 \\ 
0 & \left( 2-4\phi \right) \bar{n}_{+}^{2} & 0 & 0 \\ 
R(\bar{n}) & iS(\bar{n}) & T(\bar{n}) & iW(\bar{n}) \\ 
-iS(-\bar{n}_{+}) & R(-\bar{n}_{+}) & -iW(-\bar{n}_{+}) & T(-\bar{n}_{+})
\end{array}
\right] 
\]
\begin{equation}
\times \left[ 
\begin{array}{c}
\xi _{n}^{\alpha } \\ 
\chi _{n+1}^{\alpha } \\ 
\xi _{n}^{r} \\ 
\chi _{n+1}^{r}
\end{array}
\right] ,  \label{sec4}
\end{equation}
where $R(\bar{n})\equiv -{8}\left( 1+3\phi \right) \bar{n}^{2}+8\phi \bar{n}$%
, $S(\bar{n})\equiv 4\left( 1-\phi \right) \bar{n}^{2}+\left( 2\phi
-4\right) \bar{n}$, $T(\bar{n})\equiv -2\left( 1+8\phi \right) \bar{n}%
^{2}+4\phi \bar{n}$ , and $W(\bar{n})\equiv 2\left( 1-2\phi \right) \bar{n}%
^{2}-2\bar{n}$, and we have left out terms proportional to $y=1/\varrho $.
To calculate the corrections to the roots of (\ref{coul04}), we add (\ref
{sec4}) to (\ref{coul04}), take the determinant and equate it to zero. Here
it is necessary to evaluate the determinant using Maple, because we are
dealing with a four by four determinant. Adding matrix (\ref{sec4}) to (\ref
{coul04}) and taking the determinant we obtain, up to second order in $%
|\beta |^{2}$ 
\begin{eqnarray}
&&{256}\bar{n}^{2}(1+\bar{n})^{2}\{(\bar{n}^{4}+2\bar{n}^{3}+\left( 1+2\phi
\right) \bar{n}^{2}+2Z\phi \bar{n}+Z\phi -8Z^{2}\phi ^{2})  \nonumber \\
&&+10\phi |\beta |^{2}[\bar{n}^{4}+2\bar{n}^{3}+\left( \frac{9}{10}+\frac{Z}{%
5}+\frac{6Z\phi }{5}\right) \bar{n}^{2}-\left( \frac{1}{10}+\frac{Z}{5}+%
\frac{5Z\phi }{3}\right) \bar{n}  \nonumber \\
&&+\left( -\frac{16}{5}Z^{2}\phi ^{2}+\frac{8}{5}\phi Z^{2}+\frac{7}{5}Z\phi
+\frac{1}{2}Z\right) ]\}  \label{poly}
\end{eqnarray}
Notice that according to equation (\ref{sec4}) we still have the degenerate
roots at $\bar{n}=0$ and $\bar{n}=-1$. An inspection in (\ref{poly}) shows
that from all the roots given by (\ref{coul04}), only $\bar{n}=0$ and $\bar{n%
}=-1$ are still degenerate after the inclusion of the term in 
\mbox{$\vert$}%
$\beta |^{2}$. We will not develop it here, but for $y\neq 0$ these
degenerate roots become simple roots and each other root $0$ and $-1$ is
corrected by terms proportional to $y$%
\mbox{$\vert$}%
$\beta |^{2}$. The roots of (\ref{posiWan}) are nondegenerate and are
corrected only at second order as 
\begin{equation}
\bar{n}=-{\frac{{1}}{{2}}}\pm \sqrt{\Delta -\frac{1}{4}+\sqrt{\Delta
(9\Delta -1)}}\pm c(Z)|\beta |^{2}.  \label{corrposiWan}
\end{equation}

The function c(Z) is a complicated function of Z, and we give its value for
the most interesting values of Z, namely C(2)=0.15241, C(3)=0.11926 and
C(4)=0.10126.

As regards the second-order corrections for oscillations along the $z$
direction, the equation for $\delta z_{d}$ is changed to 
\[
{\frac{{\delta \ddot{z}_{d}}}{{\omega _{o}^{2}}}}+\delta z_{d}+{{|\beta |^{2}%
}}[(\frac{1}{2}-2\phi )\ddot{\frac{{z_{d}}}{{\omega _{o}^{2}}}}-\phi
z_{d}]=0, 
\]
and the roots are changed to 
\[
\lambda =\pm {\frac{{1}}{{2}}}[1-(\frac{1}{2}-\frac{5\phi }{4})|\beta |^{2}] 
\]
In an analogous way, we find that the second order correction to (\ref{z0})
is 
\begin{equation}
{|\beta |}^{2}\left[ 
\begin{array}{cc}
4(\frac{1}{2}-\phi )\lambda ^{2} & 0 \\ 
16(2Z\phi -\frac{1}{4}-\phi ) & -12\phi \lambda ^{2}
\end{array}
\right] \left[ 
\begin{array}{c}
c_{\alpha } \\ 
c_{r}
\end{array}
\right] .  \label{z2}
\end{equation}
Adding (\ref{z2}) to (\ref{z0}), taking the determinant and equating it to
zero, we find that the $\lambda =0$ double root is preserved and the roots
of (\ref{freq27}) are corrected to 
\begin{equation}
\lambda =\pm \sqrt{2Z\phi }(1-{\frac{{3\phi }}{{2}}}|\beta |^{2}).
\label{corfreq27}
\end{equation}

This completes the calculation of the second-order correction to all the
coulombian roots. In the next section we calculate the third-order
corrections, and table 1 shows all the roots corrected to third order.

\section{Inclusion of the Dissipative Third Order Terms}

The terms of order higher than two in (\ref{SchotP-E}) are ${\em singular}$
in the sense that they introduce the third derivative, and bring up a new
solution to the dynamics. The next natural step in the study of the
stability of the circular orbit would be to include the third-order terms in
the calculation of the eigenvalues. We can still use resonant normal form
theory in the presence of complex eigenvalues, and try to find a new
resonance condition, which would then produce a constant of the motion up to
radiative terms. We will find that many of the resonance conditions that are
satisfied by the Darwin interaction are destroyed when the third order terms
are included. For example, resonance (\ref{simplres}) is not satisfied
anymore, as one can easily check using the imaginary parts as listed in
table 1. This will not come as a surprise, and it signifies that the
radiation makes those second-order stable orbits decay. If this is the cause
of the decay, it is natural to expect that the time of decay should be of
the order of $T/(v/c)^{3}$, which is actually the case in atomic physics.
The introduction of the radiation is then seen to do two things: first, it
makes the orbit to ``dive'' to the center of the resonance island in a slow
time scale, emitting a sharp frequency. Second, since the resonance
condition is not exactly satisfyed, the electrons can escape from the island
during the long ``dive'', and decay to a lower energy state or ionize the
atom. Another cause of the decay is the nonconvergence of the resonant
normal form, whose optimal truncation produces a constant only for a finite
time scale\cite{Benettin}.

We now calculate the third-order corrections, in a way analogous to the
second order. Starting with the third-order correction to (\ref{z0}), we
find the third-order matrix 
\[
(\frac{256i\phi |\beta |^{2}}{3}\lambda ^{3})\left[ 
\begin{array}{cc}
0 & 0 \\ 
(2-Z) & {1}
\end{array}
\right] \left[ 
\begin{array}{c}
c_{\alpha } \\ 
c_{r}
\end{array}
\right] , 
\]
which describes the radiative correction to be added to (\ref{z0}), in
disregard of terms proportional to $y|\beta |^{2}$. Solving the perturbed
secular determinant, we find that the $\lambda =0$ double root is preserved
and the roots of (\ref{freq27}) are corrected to 
\begin{eqnarray}
\lambda &=&\pm \sqrt{(2Z\phi )}[1-\frac{3}{2}\phi |\beta |^{2}]  \nonumber \\
&&+\frac{64}{3}{iZ\phi }^{2}|\beta |^{3},  \label{thirdfreq27}
\end{eqnarray}
where again $i$ stands for the complex unit $i\equiv \sqrt{-1}$.

Last, let us calculate the third-order corrections to (\ref{coul0d}) and (%
\ref{coul04}). Because of symmetry, one finds that there is no third order
correction to (\ref{coul0d}). This is readily seen because the third-order
variational force is the same for both electrons, being proportional to the $%
\delta {\bf x}_{r}$ variation. The third-order correction to (\ref{coul04})
can be calculated using the third order of (\ref{SchotP-E}), in an analogous
way we used for the second order. We find the following matrix describing
the third-order correction to (\ref{coul04}) 
\[
\frac{128}{3}i\phi |\beta |^{3}\left[ 
\begin{array}{cccc}
0 & 0 & 0 & 0 \\ 
0 & 0 & 0 & 0 \\ 
2(2-Z)\bar{n}^{3} & 0 & 2\bar{n}^{3} & 0 \\ 
2(2-Z)\bar{n}_{+}^{3} & 0 & 2n_{+}^{3} & 0
\end{array}
\right] \times 
\]
\[
\left[ 
\begin{array}{c}
\xi _{n}^{\alpha } \\ 
\chi _{n+1}^{\alpha } \\ 
\xi _{n}^{c} \\ 
\chi _{n+1}^{c}
\end{array}
\right] . 
\]
Adding the above matrix to (\ref{coul04}), add the second-order (\ref{sec4})
and take the determinant. The result produces a correction to equation (\ref
{poly}) given by 
\begin{eqnarray}
\delta _{3} &=&-{\frac{16384{|\beta |^{3}i}}{{3}}}(1+\bar{n})^{2}\bar{n}%
^{2}(1+2\bar{n})\times  \nonumber \\
&&(\bar{n}^{4}+2\bar{n}^{3}+\bar{n}^{2}+Z\phi \bar{n}^{2}+Z\phi n+Z\phi ).
\label{third04}
\end{eqnarray}
From the above, it is easy to see that the eigenvalues $\bar{n}=-1/2$, $\bar{%
n}=-1$ and $\bar{n}=0$ do not acquire any imaginary part at third order. It
is also straightforward to add (\ref{third04}) as a perturbation to (\ref
{coul04}) and calculate the imaginary correction to (\ref{posiWan}) (using
the program Maple). The imaginary correction is of type id(Z), and the most
interesting values of d are d(2)=0.766, d(3)=0.740 and d(4)=0.731. In table
1 we show all the corrections of the eigenvalues, including order 
\mbox{$\vert$}%
$\beta |^{3}$ correction. We only showed in the table 1 the correction to
the regular roots, by which we mean the ones that were already roots of the
Coulomb dynamics. The introduction of the third order brings up nine more
singular roots, which we do not consider. Last, we did not show in table 1
the unstable pair of eigenvalues (\ref{negWan}) of the radial instability
either, which are part of the eighteen coulombian eigenvalues but are not
interesting for resonance conditions. The unstable mode should not take part
of a resonant condition, but of course it will be part of the next orders of
the normal form.

\section{Some special orbits resulting from the Resonance condition}

In this section we explore the orders of magnitude of some of the orbits. A
rigorous approach to the material in this section would be to study the
nonlinear stability first. This very hard task is not done yet and in
appendix C we outline what remains to be done. The section is designed to
inspect some of the stable orbits predicted by resonant normal form theory
using the information already at hand. The guiding dynamical principle we
use is that resonances with the minimal integer multipliers are the most
important. It is known in general \cite{Lichtenberg} that the size of the
resonance islands varies as $\exp (-o),$ where o is the order of the
resonance. This solves the paradox as to the infinite number of possible
resonances: the ones with a high order occupy an exponentially small area of
phase space, which makes them very unlikely. The situation is analogous to
quantum mechanics, where there is always an infinity of energy levels.
Nevertheless, in practice only a very small finite number of them exists in
experimental situations. For example in the Balmer series, only the first
twelve frequencies of the series can be observed as emission lines in very
diluted gaseous states\cite{Trabert,Bohr}. Since very high quantum states
are too extended in space, one needs very rarefied gases and large
astronomical masses of gas to produce a measurable signal. We consider in
this section only some of the resonances with minimal ordering. The
comparisom with the quantum results is very illuminating because it
immediately suggests what further results are worthed pursuing with
nonlinear dynamics techniques. It also brings up some of the strengths
(surprisingly good agreement), and weaknesses (lack of a selection rule so
far) of the research so far. In principle the dynamical system with
retardation and self-interaction is a nonlinear system that can emit sharp
lines, but it does not have to agree quantitatively with the quantum
results. It is very intersting that it does so to some extent. We have by no
means exhausted the material that would fit in this section.

An approximation that we use in examining the resonance conditions is that
we only include the correction to the frequency that comes with linear
order, because this is the largest correction for small $|\beta |.$ Let us
now consider the special case of helium: As we mentioned before, according
to the Darwin Lagrangian, helium has always an angular momentum-like
constant of motion and an energy-like constant. In the neighborhood of some
select circular orbits, another complex analytic constant might exist, which
could make those orbits nonlinearly stable. The constant we find here
involves the amplitude of the normal mode corresponding to (\ref{freq27}) in
a combination with normal modes along the plane. This is possible because
these modes are coupled at higher orders in the oscillation amplitude. We do
not want to involve the linear modes describing the circular instability in
the first resonant term of the constant, but the amplitudes of the unstable
mode will naturally appear in the higher monomials of the series for the
constant. By the way, this is how the circular instability can be
equilibrated inside the small resonance island defined by the constant.
Besides, the circular pair (\ref{negWan}) are complex eigenvalues and would
not satisfy a simple resonance condition in combination with the other real
eigenvalues. (one would have to vanish the real and imaginary parts of the
resonance condition separately, which would require two integers at the
best.)

The circular orbit is a fixed point of the autonomous vector field
describing the dynamics in a system rotating with the frequency of the
circular orbit. To develop the normal form about the fixed point, we must
move to this rotating frame. We recall that $\bar{n}$ is a frequency of
oscillation, in units of $2\omega $, of the variational dynamics, according
to (\ref{Floquet}). In the rotating system, the new frequencies for
oscillation along the plane are found by adding $1/2$ to the formulas (\ref
{posiWan}) and (\ref{bifline}). The frequency (\ref{freq27}) for the $z$%
-oscillation is unchanged. A new resonance condition involving the root (\ref
{bifline}) is the easiest to be satisfied for small values of 
\mbox{$\vert$}%
$\beta |$. Therefore, we suggest to look for a resonance among the
frequencies of (\ref{posiWan}), (\ref{freq27}), and (\ref{bifline}), which
we rename as $\omega _{1}$, $\omega _{2}$, and $\omega _{3}$, respectively,
in the rotating frame, and which in the case of helium evaluate to 
\begin{eqnarray}
\omega _{1} &\equiv &\sqrt{\frac{{2}}{{7}}}\simeq 0.5345\ldots  \nonumber \\
\omega _{2} &\equiv &\sqrt{\frac{{3+\sqrt{3}2}}{{28}}}\simeq 0.5560\ldots 
\nonumber \\
\omega _{3} &\equiv &\sqrt{\frac{{12}}{{7}}}|\beta |\simeq 1.3093|\beta |,
\label{resonants}
\end{eqnarray}
in units of $2\omega $, and we have disregarded corrections proportional to $%
y$ and 
\mbox{$\vert$}%
$\beta |^{2}$. According to standard normal form theory, the necessary
condition to have an additional analytic \cite{Gustav,Furta} constant of the
motion in the neighborhood of a fixed point is a resonance among the
frequencies. By inspection, we find that a new quartic resonance involving
the above three frequencies and with the minimal integer multipliers is of
type 
\begin{equation}
\omega _{1}-\omega _{2}+2\omega _{3}=0.  \label{simplres}
\end{equation}
This resonance is satisfyed for 
\mbox{$\vert$}%
$\beta |$ given by 
\[
|\beta |=0.0082\ldots . 
\]
The Coulombian binding energy of a circular orbit in a two-electron atom can
be written as $E=-mc^{2}|\beta |^{2}$. This energy, in atomic units, for the
above value of 
\mbox{$\vert$}%
$\beta |$ is 
\[
E=-1.265a.u. 
\]
Of course, other resonance conditions are possible: for example, we could
put an integer number in the resonance condition, as in 
\begin{equation}
\omega _{1}-\omega _{2}+2n\omega _{3}=0,  \label{nres}
\end{equation}
and the corresponding values of 
\mbox{$\vert$}%
$\beta |$ are given by 
\[
|\beta |={\frac{{0.0082}}{{n}}}. 
\]
The discrete circular orbits corresponding to this have binding energies
given by 
\[
E=-{\frac{{1.265}}{{n^{2}}}}a.u. 
\]
In the next section we show how to construct the extra complex constant of
the motion in a Maclaurin series (resonant normal form) when condition (\ref
{nres}) is satisfied.

The surprising fact that we were able to pick particular circular orbits is
a genuine signature of the nonlinear dynamics, because the linear
eigenvalues depend on the circular orbit through the parameter 
\mbox{$\vert$}%
$\beta |$. The extra constant of motion, together with the four other
constants of helium can stabilize the orbit by returning the unstable
direction back to the neighborhood of the special orbit. This will happen
only inside some resonance region\cite{deluca}. We discuss this further in
appendix C. The frequencies of the resonant orbits satisfying (\ref{nres})
are given by 
\[
\omega ={\frac{{0.8101}}{{n^{3}}}}a.u., 
\]
and the frequency of the z-oscillation in the stable manifold of the orbit
can be obtained by multiplying the above frequency by $2\sqrt{2/7}$, as of (%
\ref{freq27}) 
\begin{equation}
w_{z}={\frac{{0.866}}{{n^{3}}}}.  \label{stabz}
\end{equation}

Supposing that the dynamics in the neighborhood of the resonant orbits is
stable , one could expect that the stable oscillations about this orbit
could emit a sharp line. The only condition to emit a sharp line is to
oscillate with the same frequency for a long enough time (of the order of
the inverse of the width of the line). This condition is fulfilled because
of the finite time stability of the resonant orbit. The correct frequency of
these stable oscillations can only be obtained after the nonlinear stability
is performed, by linearizing about the center of the resonance islands. Here
we will assume that the frequencies of \ref{stabz} are an approximation to
those.

For the circular orbit corresponding to (\ref{stabz}) with $n=1$, the
frequency of the z-oscillation in the stable manifold is 0.7956 atomic
units. The transition from the first excited state of parahelium to the
ground state ($2^{1}P\rightarrow 1^{1}S$) corresponds to a frequency of $%
2.9037-2.1237=0.7799$ atomic units\cite{Chen}, which is a $9\%$ difference.
For $n=2$, (\ref{stabz}) evaluates to 0.1083 atomic units, and the frequency
for the transition ($3^{1}P\rightarrow 2^{1}S$) in parahelium is $%
2.1459-2.0551=0.0908$ atomic units, which is again a $9\%$ difference. Last,
the asymptotic form of the quantum energy levels of helium\cite{Bethe}, both
parahelium or orthohelium, is 
\[
E={\frac{{-Z^{2}}}{{2}}}-{\frac{{(Z-1)^{2}}}{{2(n_{r}+L+1)^{2}}}}, 
\]
with $Z=2$, which is a first approximation to the Rydberg-Ritz spectroscopic
term\cite{Trabert}. The frequency of the line emitted by transitioning to
the neighboring level, calculated from the above formula with $\Delta L=1$
is approximated by 
\[
w={\frac{{1}}{{(n_{r}+L+1)^{3}}}}. 
\]

This above equation is a quantum formula which we write just to compare with
(\ref{stabz}). Notice that (\ref{stabz}) agrees with it to within $13\%$.

\smallskip We can also produce an estimate for the width of the line as
follows: The third-order correction to the frequency of (\ref{thirdfreq27}) $%
(w_{z})$ is imaginary and with the same sign for the two values of $\lambda $%
. Therefore the $z$-oscillation is stable and decays with a coefficient
which, according to (\ref{deflamb}), is the imaginary part of $2\omega
\lambda $. For example in the case of helium, $Z=2$, and $\phi =1/14$ , this
imaginary correction to (\ref{stabz}) evaluates to 
\[
\nu =i\frac{64}{49}|\beta |^{3}\omega , 
\]
and this oscillation is then part of the stable manifold of the circular
orbit. Along this decaying oscillation, the perturbed circular orbit decays
back to the perfectly circular special circular orbit. The radiative
self-interaction has already been shown to produce good approximations for
the linewidth in other situations\cite{Seidl}. If we evaluate the above
result for the line at $\omega =0.7959a.u.$, we find that it is 1.5 times
the experimental linewidth for this transition\cite{Chen,exper}. We stress
again that this is not exactly the probability of decay of the orbit: As we
already mentioned, the resonance condition (\ref{simplres}) is not satisfyed
with the inclusion of the third-order terms, which implies that the constant
of motion is destroyed. Because of this, the perturbed orbit can decay not
only back to the perfectly circular circular orbit, but also to the
lower-energy ground state. The dynamical linewidth would be the full
probability to escape from the atracting resonance region of the stable
orbit, and the calculation of this is beyond the scope of the present paper,
but since the damping of the oscillation is causing the decay, one would
expect a number of the order of this damping for the inverse of the
linewidth, which is again a good agreement, since we found 1.5 times the
correct quantum result.

\smallskip Let us briefly consider the case of the $Li^{+}$ ion: The
frequencies of \ref{resonants} can be easily recalculated for the case of $%
Li^{+}$ by using the material of sections III and IV with $Z=3$ and $\phi
=1/22$ .We find\smallskip 
\begin{eqnarray}
\omega _{1} &\equiv &\sqrt{\frac{3}{11}}\simeq 0.5222\ldots  \nonumber \\
\omega _{2} &\equiv &\sqrt{\frac{5{+}\sqrt{60}}{{44}}}\simeq 0.5382\ldots 
\nonumber \\
\omega _{3} &\equiv &\sqrt{\frac{{12}}{{7}}}|\beta |\simeq 1.2792|\beta |,
\end{eqnarray}

and a simple resonance condition like

\begin{equation}
\omega _{1}-\omega _{2}+\omega _{3}=0.
\end{equation}

will determine the value of $|\beta |$ to be

\[
|\beta |=0.0125. 
\]
The stable orbit has a Coulombian energy of

\[
E=-2.932a.u. 
\]
and the frequency of the z-oscillation mode is

\[
w_{z}=1.97a.u. 
\]
again in very good agreement with quantum mechanics. The first two quantum
energies of para-lithium $Li^{+}$ \cite{Seaton} are: ground state, $%
1^{1}S:E=-7.278a.u.$ and $2^{1}P:E=-5.30a.u.$\smallskip The frequency of the
dipole transition becomes $w=7.278-5.3=1.978a.u.,$ in very good agreement
with our above value of $w_{z}$ (within one percent). Notine that this time
we started the resonance condition with n=1 instead of n=2 as for helium. It
would be futile to go on without knowing the reason for the special integer
combination. In appendix C we show that this selection rule requires the
knowledge of the next term of the resonant normal form . This would be also
the case if we tryed to apply this to find some states of $H^{-}(Z=1)$ : It
is known that H-minus has very few, if any excited states, and accordingly,
the nonlinear stability should predict that all most of the circular orbits
are unstable. The frequencies we obtained are in good agreement with the
sharp lines of helium. The values of the energies do not agree so well with
the quantum energies, but are nevertheless in the same order of magnitude.
This comparisom should be repeated after one knows more about this dynamical
system.

\section{Discussions and Conclusion}

\label{discussion}

Historically (1912), coulombian many-electron atoms (``saturnian atoms'')
were investigated by astronomers\cite{Nicholson} prior to Bohr, who was well
aware of these studies. For oscillations perpendicular to the plane of the
orbit, the Coulomb dynamics is stable and the ${\em ratio}$ of many lines
obtained by Nicholson for perpendicular oscillations agreed with the spectra
of the Orion nebula and the Solar corona\cite{Nicholson}. Of course
Nicholson was assuming a special radius for the orbits, which he did not
know how to calculate, and this radius would disappear when one took the
ratio of two lines of the stable manifold of an orbit (because of a
degeneracy of the Coulomb interaction). Bohr was originally favorable to the
use of ordinary mechanics to describe those stable perpendicular oscillations%
\cite{Bohr}. For oscillations along the plane of the orbit, Nicholson first
found the now well-studied Coulombian radial instability\cite
{Poirier,Wintgen}, which was then a hindrance for the theory\cite{Bohr}. It
is of historical interest to stress that it was the radial instability that
first motivated Bohr to postulate a discrete set of special, more stable
orbits, which proved to be a very fruitful intuition\cite{Bohr}. The
original critic of Bohr\cite{Bohr} to Nicholson\cite{Nicholson}, was that
the circular instability would make the atoms too ``fragile'' to
disintegration, and unable to emit a sharp line. This led Bohr to conjecture
that, along some special stable orbits, some ``quantum'' mechanism would
supersede normal mechanics and prevent the radial instability\cite{Bohr}.

As we mentioned in the introduction, the classical chaotic Coulomb dynamics
of helium has been widely studied in connection with the recent interest in
periodic-orbit quantization\cite{Wintgen,Kaneko}. The circular instability
is now known to lead to self-ionization, after a long-term chaotic
transient, for most initial conditions\cite{Kaneko}. Of course, those
studies are for a different dynamical system, with lesser constants of
motion, once the extra constant needs the retardation to exist. It would be
of interest to repeat these numerical studies with the inclusion of
retardation.

As regards to where the circular orbits ultimately decay to, we conjecture
that the lowest energy bound-state in helium could be a ``non-mechanical''
orbit of symmetric collinear motion. This orbit has zero angular momentum
and zero electric dipole moment. The avoided three body collision can be
provided by a singular mechanism analogous to the one of Eliezer's theorem
for hydrogen\cite{Eliezer,Parrott}. In hydrogen, Eliezer's theorem predicts
that the electron will runaway from the atom, which is closely related to
the fact that one has always dipole radiation in atomic hydrogen. The
symmetric collinear motion in helium is not radiating in dipole, and
therefore one could expect a physical solution, differently from the case of
hydrogen\cite{Parrott}. It is intersting to compare these
zero-angular-momentum singular solutions of Eliezer's theorem to the
divergent series of the Lamb shift\cite{Bethe,Trabert} in quantum mechanics.
The Lamb shift appears in quantum mechanics because of a singular
interaction with the electromagnetic field and is only pronounced for states
of zero angular momentum\cite{Bethe,Trabert}. In the dynamical approach, the
solutions with zero angular momentum and zero dipole will be non-mechanical
collinear orbits that get very close to the nucleus. This in turn causes the
Page series to diverge or be asymptotic at the best.

A recent use of resonances in perturbation theory worth mentioning here was
on the problem of the time of stability of integrable tori. Here one
exploits the resonances among the unperturbed frequencies\cite
{Benettin,Lochak}. Those results go under the name of Nekhoroshev bounds for
Arnold Diffusion. Simply stated, the results say that the actions of an $%
\epsilon $-perturbed Hamiltonian system are kept approximately constant for
a time of the order of 
\[
T\simeq ({\frac{{1}}{{\epsilon }}})\exp {\ ([{\frac{{1}}{{\epsilon }}}]^{a})}%
, 
\]
where $a\equiv (1/d)$ and $d$ is the maximal number of unperturbed
frequencies linearly independent over the rationals\cite{Galgani,Lochak}.
Every time there is a resonance among the frequencies, $d$ is reduced and
the torus has an exponentially longer time of stability. This phenomenon is
named ``stability by resonance''\cite{Lochak}. In connection with these
modern Arnold Diffusion results, it is interesting to mention that they
shone new light onto the old problem of the ultra-violet catastrophe, which
was the historical motivation for Plank's hypothesis\cite{Galgani,Nature}.
Today, also many numerical results exist showing that a set of coupled
oscillators might never reach equipartition\cite{ddl}, the reason being a
super-slow Arnold Diffusion.

As regards further research to be done, we have only been able to study the
circular stationary orbits, and it would be of much interest to study the
most general elliptical stationary orbits and associated energies and
spectra, which should present a richer structure. Along elliptical orbits
one can still use regular perturbation theory, in the same way used for
circular orbits here, but now the parametric problem associated is more
complicated. It might be that the use of the Kustanheimo transformation\cite
{Stiefel} to regularized coordinates will simplify the problem of elliptical
orbits. For zero-angular-momentum collinear orbits, it might be necessary to
introduce the retardation effects in a non-perturbative way. To ``quantize''
those zero-angular-momentum orbits using resonant normal form theory is an
open problem of much theoretical interest.

One would expect some discussion about spin: Notice that our simplified
dynamical system is based on classical pointlike charges with no spin. Of
course, we could include spin in the dynamics in a phenomenological way,
similar to the usual way it is introduced in quantum mechanics. We have not
done that yet. Second, quantum mechanical spin is a relativistic effect and
introduces a correction comparable to the second-order retardation correction%
\cite{Bethe,Trabert}. As a matter of fact, the Breit operator for helium is
actually produced starting from the Darwin Lagrangian\cite{Bethe,Trabert}.
We believe that a correct discussion of spin issues can only be made after
we know more details about our dynamics, such as the frequency correction
due to motion along resonance islands and etc.

As a summary, we presented a complete account of the linear stability of a
two-electron atom along circular orbits in the presence of retardation and
self-interaction. We calculated all the linear eigenvalues up to third order
in $(v/c)$. We considered the necessary condition for an extra constant and
showed how to construct this extra analytic constant. We have also shown
that electrodynamics assigns a dynamical system to helium, which can be
responsible for the emission of sharp spectral lines. (The sharp line is
emitted by ``diving'' to the center of the resonance island, radiating out
some energy and decaying to a lower state). We found that the frequency of
the sharp line associated with the lowest stable orbits agrees with the
highest frequency in the spectrum of helium to within $8\%$. We do not know
of prior results on the existence of these stable electromagnetic orbits,
which appear as a genuine effect of the nonlinear dynamics prescribed by
Maxwell's electrodynamics to atomic physics. We have barely touched the
study of this dynamical system, and much research remains to be done,
specially in the nonlinear stability along the lines of appendix C.

\appendix

\section{Variation of the Coulomb Interaction}

\label{appA}

In this Appendix we calculate the variation of the terms appearing in the
Page series, equation (\ref{SchotP-E}). To obtain the variation of the
Coulomb force along the plane of the orbit, one has to evaluate 
\begin{equation}
\delta ({\frac{{\bf x}}{{|{\bf x}|^{3}}}})={\frac{{\delta {\bf x}}}{{r^{3}}}}%
-{\frac{{3(x\delta x+y\delta y){\bf x}}}{{r^{5}}}},  \label{eqdx}
\end{equation}
where $r=|{\bf x}|$ and ($\delta x$, $\delta y$) are the functions of time
representing the variation about the circular orbit 
\begin{equation}
x=\pm r\cos (\omega t),\mbox{  }y=\pm r\sin (\omega t).  \label{eqorb}
\end{equation}
Substituting (\ref{eqorb}) into (\ref{eqdx}) we obtain 
\begin{eqnarray}
\delta ({\frac{{x}}{{r^{3}}}})_{n} &=&\{\delta x-{\frac{{3}}{{2}}}(1+\cos
(2\omega t))\delta x-{\frac{{3}}{{2}}}\sin (2\omega t)\delta y\}_{n}, 
\nonumber \\
\delta ({\frac{{y}}{{r^{3}}}})_{n} &=&\{\delta y-{\frac{{3}}{{2}}}(1-\cos
(2\omega t))\delta y-{\frac{{3}}{{2}}}\sin (2\omega t)\delta x\}_{n},
\end{eqnarray}
where the subscript $n$ indicates the $nth$ Floquet component. Using
equation (\ref{Floquet}) we find $(\delta x)_{n}=x_{n}$ and $(\delta
y)_{n}=y_{n}$ and 
\begin{eqnarray}
\delta ({\frac{{x}}{{r^{3}}}})_{n} &=&-{\frac{{1}}{{2r^{3}}}}(x_{n}+3U_{n}),
\nonumber \\
\delta ({\frac{{y}}{{r^{3}}}})_{n} &=&-{\frac{{1}}{{2r^{3}}}}(x_{n}+3V_{n}),
\label{vacoul}
\end{eqnarray}
where we have used $U_{n}$ and $V_{n}$ as defined by equation (\ref{defUL}).
An economic way to write this equation is 
\[
\delta ({\frac{{\bf x}}{{r^{3}}}})_{n}=-{\frac{{1}}{{2r^{3}}}}({\bf x}_{n}+3%
{\bf K}_{n}). 
\]
When we consider the second-order terms of the Schott-Page force and the
relativistic correction of the mass, we need to evaluate the variation of
many other functions besides $x/r^{3}$. This is done in Appendix \ref{appB},
in a way analogous to the above calculations. Last, along the $z$ direction
the variation is simply 
\[
\delta ({\frac{{z}}{{r^{3}}}})={\frac{{\delta z}}{{r^{3}}}}. 
\]

\section{Variation of the Second-Order Forces}

\label{appB} In this Appendix we consider the terms of order $(v/c)^{2}$ in
the force. One contribution comes from the first relativistic correction to
the electronic masses (the $\alpha $ particle is at rest). The $x$ component
of this term is 
\[
\delta M_{e}a_{x}^{(2)}={\frac{{M_{e}}}{{c^{2}}}}\delta (\dot{{\bf x}}\cdot 
\ddot{{\bf x}})\dot{x}_{e}+{\frac{{M_{e}}}{{2c^{2}}}}\delta (|{\bf x}%
_{e}|^{2}\ddot{x}_{e}) 
\]
Substituting the equation for a circular orbit, we find 
\begin{eqnarray*}
(\delta M_{e}a_{x}^{(2)}) &=&{\frac{{M_{e}\omega ^{2}}}{{2}}}|\beta
|^{2}[(2-\cos (2\omega t)){\frac{{\delta \ddot{x}_{e}}}{{\omega ^{2}}}} \\
&&-\sin (2\omega t){\frac{{\delta \ddot{y}_{e}}}{{\omega ^{2}}}}+2(\sin
(2\omega t){\frac{{\delta \dot{x}_{e}}}{{\omega }}}-\cos (2\omega t){\frac{{%
\delta \dot{y}_{e}}}{{\omega }}})]
\end{eqnarray*}
Using the above equation, equation (\ref{Floquet}) for $\delta {\bf x}$ and
definition (\ref{defUL}) for $U_{n}$ and $L_{n}$ we obtain the Floquet
component 
\[
(\delta M_{e}a_{x}^{(2)})_{n}=M_{e}\omega ^{2}|\beta _{e}|^{2}(-4\bar{n}%
^{2}x_{n}^{e}+2\bar{n}^{2}U_{n}^{e}+2in\bar{V}^{e}) 
\]
The $y$ component can be calculated in an analogous way. To write the two
components in a concise vector form we define an extra useful vector
quantity 
\[
\tilde{{\bf K}}_{n}^{e}=\left[ 
\begin{array}{c}
V_{n} \\ 
-U_{n}
\end{array}
\right] , 
\]
This quantity is obtained from the vector ${\bf K}$ (defined just bellow
equation (\ref{defUL}) by exchanging the components and changing the sign of
the second component). Making use of the above, we write the relativistic
second-order force as 
\[
(\delta M_{e}{\bf a}^{(2)})_{n}=M_{e}\omega ^{2}|\beta _{e}|^{2}(-4\bar{n}%
^{2}{\bf x}_{n}^{e}+2\bar{n}^{2}{\bf K}_{n}^{e}+2i\bar{n}\tilde{{\bf K}}%
^{e}) 
\]
Using a procedure analogous to the above, we can obtain the variation of the
second-order electric interaction of (\ref{SchotP-E}). First, the variation
of the second-order electric field produced by the electron at the point $p$
on the observation point $o$ is calculated. The vector pointing from the
electron to the point $p$ (where another particle is) is ${\bf x}={\bf x}%
_{k}-{\bf x}_{e}$ ($\kappa $ designates the other particle). In all cases of
interest we have ${\bf x}=-g{\bf x}_{e}$. It is easy to verify that for
electron-proton interaction $g=1$ and for electron-electron interaction $g=2$%
. According to (\ref{SchotP-E}), the variation of the electric field is
proportional to the product of the charges times 
\begin{eqnarray}
\delta ({\frac{{|\beta _{e}|^{2}{\bf x}}}{{2r^{3}}}}-{\frac{{\dot{{\bf \beta 
}}_{e}}}{{2rc}}}-{\frac{{({\bf n}\cdot \dot{\beta}_{e}){\bf n}}}{{2rc^{2}}}}%
)_{n} &=&  \nonumber \\
{\frac{{|\beta _{e}|^{2}}}{{2r^{3}}}}\{({\frac{{g-1}}{{2}}})({\bf x}_{n}+3%
{\bf K}_{n})+2\bar{n}^{2}g^{2}({\bf x}_{n}^{e}+{\bf K}_{n}^{e} &&)+ 
\nonumber \\
4\bar{n}^{2}({\frac{{r}}{{R}}})^{2}{\bf x}_{n}^{e}-2ig\bar{n}(\tilde{{\bf x}}%
_{n}^{e}-\tilde{{\bf K}}_{n}^{e}) &&\},
\end{eqnarray}
where $r$ is the interparticle distance and $R$ the radius of the electronic
orbits. The tilde over ${\bf x}$ means the vector obtained from ${\bf x}$ by
exchanging the components and changing the sign of the second component. 
\[
\tilde{{\bf x}}_{n}^{\kappa }=\left[ 
\begin{array}{c}
y_{n}^{\kappa } \\ 
-x_{n}^{\kappa }
\end{array}
\right] , 
\]
where again the superscript $\kappa $ can take the values 1,2,$\alpha $,$r$
and $d$, to represent a particle's coordinate or a combination of the
coordinates, as defined in (\ref{defradrel}). The second-order electric
field produced by the nucleus on each electron is 
\[
-{\frac{{2e^{2}|\beta |^{2}\bar{n}^{2}}}{{R^{3}}}}(3{\bf x}_{n}^{\alpha }+%
{\bf K}_{n}^{\alpha }), 
\]
the same for the two electrons. Last, we calculate the variation of the
second-order magnetic forces. The $\alpha $ particle is at rest and does not
produce any magnetic field. The two electrons produce a magnetic field at
the center of the orbit, and the Lorentz magnetic force over the nucleus is 
\[
(\delta {\bf F}_{M}^{\alpha })_{n}=-{\frac{{4Zie^{2}\omega l_{z}}}{{%
M_{e}c^{2}R^{3}}}}\bar{n}\tilde{{\bf x}}_{n}^{\alpha }, 
\]
where $l_{z}$ is the $z$-component of the angular momentum of one electron.
The variation of the electron-electron magnetic force done by electron 2
over electron 1 is 
\begin{eqnarray}
(\delta {\bf F}_{21})_{n} &=&{\frac{{e^{2}|\beta _{e}|^{2}}}{{r^{3}}}}\{2i%
\bar{n}(2\tilde{{\bf x}}_{n}^{e1}-\tilde{{\bf x}}_{n}^{e2}  \nonumber \\
&&+\tilde{{\bf K}}_{n}^{e2})-({\bf x}_{n}^{d}+{\bf K}_{n}^{d})\}.
\end{eqnarray}
For the force done by electron 1 over electron 2 just interchange the
indices and the sign in front of ${\bf x}_{n}^{d}$ and ${\bf K}_{n}^{d}$.

\section{Construction of the Extra Complex Constant of Motion}

\label{construction}

To construct the resonant normal form, it is again convenient to move to the
rotating system. For this part we disregard the acceleration of the nucleus,
so that ${\bf x}_{\alpha }=0$. It is also convenient to use the Hamiltonian
formalism associated with the Darwin Lagrangian, written in terms of the
coordinates ${\bf x}_{r}$ and ${\bf x}_{d}$, as defined by equation (\ref
{defradrel}), and the conjugate momenta ${\bf p}_{r}$ and ${\bf p}_{d}$. We
make a last transformation to the normal mode coordinates of the Jacobian
matrix about the fixed point corresponding to the circular orbit. Let $u_{i}$%
, $i=1,\ldots ,12$, be the coordinates corresponding to the normal modes of
frequencies $\omega _{i}$, $i=1,\ldots ,12$, respectively. These coordinates
will be complex-valued linear functions of the coordinates and momenta.
Because of the simplectic symmetry, the eigenvalues exist in pairs $\omega $
and $-\omega $\cite{Lichtenberg}. We order the modes such that $u_{12-i}$ is
the coordinate corresponding to the frequency $\omega _{12-i}=-\omega _{i}$.
Because there are six coordinates and six conjugate momenta, we have twelve
normal coordinates. For example, the normal mode solution for the coordinate 
$u_{1}$ will be given by 
\[
u_{1}(t)=u_{1}(0)\exp (i\omega _{1}t), 
\]
and so on for the other coordinates. We define these coordinates to describe
the linearization about the fixed point corresponding to the circular orbit,
so that at the circular orbit all coordinates are zero.

We want to find an extra constant with an analytic form, which implies it
has a formal Taylor series about the fixed point (not necessarily
convergent). For instance, the integer powers appearing in this Taylor
series are the cause of the integer numbers in the necessary resonance
condition\cite{Furta}. Before we go on, let us introduce some definitions to
simplify the exposition: Because we consider a Maclaurin expansion of the
constant of motion, some notation concerning monomials is in place: We write
a monomial in the coordinates $u_{j}$, $j=1,\ldots 12$ as 
\[
U^{{\bf k}}\equiv u_{1}^{k_{1}}u_{2}^{k_{2}}\ldots u_{12}^{k_{12}}, 
\]
where ${\bf k}$ is a twelve-dimensional vector of integer components. For
example, if the $ith$ coordinate is not present in the monomial, then $%
k_{i}=0$. We use the convention that when $k_{i}$ is negative, the
coordinate to be used is the complement of $k_{i}$ to 12 with the positive
power ($k_{12-i}=-k_{i}$). For example 
\[
u_{1}(t)^{-1}\equiv u_{7}(t), 
\]
and so on for all the other coordinates. Mode 7, by definition, has a
frequency that is the negative of $\omega _{1}$. With this convection, even
though we are working with negative integers, all the powers of the
coordinates in the series for the constant are positive, as it should for
any good Taylor series.

Let us define the order of a monomial by 
\[
o=|k_{1}|+|k_{2}|+\ldots +|k_{12}|,
\]
and the frequency associated with a monomial as 
\[
\Delta \omega _{{\bf k}}=k_{1}\omega _{1}+k_{2}\omega _{2}+\ldots
+k_{12}\omega _{12}.
\]

The necessary condition for the existence of the extra analytic constant is
that some resonance condition must be satisfied among the linear
eigenvalues. In the following we show how to construct the constant starting
from a leading resonance\cite{whittaker,Contopoulos,Gustav,Furta}. Let us
assume that a resonance defined by $k_{o}$ exists 
\begin{equation}
\Delta \omega _{{\bf k_{o}}}=0.  \label{rezero}
\end{equation}
Our proposed analytic constant must have a Maclaurin series expansion given
by 
\[
C=\sum_{{\bf k}}C_{{\bf k}}U^{{\bf k}},
\]
where the $C_{{\bf k}}$ are constant numbers. This is an analytic function
by construction, because it involves only integer powers of the coordinates.
We now construct a constant where the lowest-order monomial is defined by $%
{\bf k}_{o}$ as of (\ref{rezero}), plus higher-order monomials necessary to
vanish the time derivative at higher orders. For example, if the resonance
condition (\ref{simplres}) is our leading resonance, then we can construct a
nontrivial analytic constant by starting the Maclaurin series with $k_{1}=1$%
, $k_{8}=1$, $k_{3}=2$, and $k_{2}=k_{3}=\ldots =k_{12}=0$, as of (\ref
{simplres}). To produce the successive powers of the constant, let us write
the constant as 
\begin{equation}
C=U^{{\bf k}_{0}}+\sum_{{\bf k}_{1}}C_{{\bf k}_{1}}U^{{\bf k}_{1}}+\ldots 
\label{constant}
\end{equation}
where we have separated out the contribution of the leading term at ${\bf k}%
_{0}$, the contribution of the monomials of immediately next order, which we
index with ${\bf k}_{1}$, and the dots represent monomials of higher than
immediately next orders. This is a complex valued function, which means two
real functions of phase space. Of course, starting the series with the
lattice vector $-{\bf k}_{0}$ will produce the complex conjugate of the same
function. Next we evaluate the time derivative of (\ref{constant}).

The time derivative of the leading monomial produces the same monomial
multiplied by $\Delta \omega _{{\bf k}_{0}}=0$, plus higher-order terms, as 
\[
{\frac{{d}}{{dt}}}(U^{{\bf k}_{0}})=\Delta \omega _{{\bf k}_{0}}U^{{\bf k}%
_{0}}+\sum_{{\bf k}_{1}}N_{{\bf k}_{1}}U^{{\bf k}_{1}}+\ldots 
\]
This monomial is then almost a constant, up to higher order monomials. To
produce the higher monomials of the constant, let us focus on one such term
of immediately next order, corresponding to a vector ${\bf k}_{1}$ and with
coefficient $N_{{\bf k}_{1}}$. The time derivative of $C$ is 
\begin{eqnarray}
{\frac{{dC}}{{dt}}} &=&\Delta \omega _{{\bf k}_{0}}C_{{\bf k}_{0}}U^{{\bf k}%
_{0}}+\sum_{{\bf k}_{1}}N_{{\bf k}_{1}}U^{{\bf k}_{1}}+\ldots  \nonumber \\
&&+\sum_{{\bf k}_{1}}\Delta \omega _{{\bf k}_{1}}C_{{\bf k}_{1}}U^{{\bf k}%
_{1}}+\ldots
\end{eqnarray}
It is easy to see that the next order nonlinear contribution of the leading
monomial to the time derivative can be canceled by choosing 
\[
C_{{\bf k}_{1}}=-N_{{\bf k}_{1}}/\Delta \omega _{{\bf k}_{1}}. 
\]

It is well known that this perturbation scheme does not produce a convergent
constant, which is not a problem if we are investigating stability for a
finite time scale. The usual procedure is to terminate the series using some
optimal truncation\cite{Benettin}. This produces a quasi-constant for a long
time-scale, which still provides stability in the finite time for the
special orbits. The detailed investigation of this issue is beyond the scope
of the present paper. Here we are mainly concerned with exploring the
necessary condition for the existence of such a constant. Just for
illustration, let us assume that optimal truncation means considering the
first term of the series plus the largest next order term. As this term
contains resonances, the largest next order term will come divided by the
smallest resonance, which in our case is proportional to $|\beta |^{2}.$ To
see what kind of constraint this constant can provide, let us assume that
the perturbations about the orbit have a magnitude of size $\zeta .$ The
constant expressed in terms of $\zeta ,$ with the first and most resonant
next term included is 
\[
C=\zeta ^{2+2n}-\rho \frac{\zeta ^{4+2n}}{|\beta |^{2}}+...
\]

In the above equation, the first power is because the order of the resonance
in assumed to be $2+2n,$ and we are also assuming that the next most
important term is of order $4+2n$. The constant $\rho $ will be a function
of $n$ and $Z$, and the selection rule will be $\rho \left( n,Z\right) >0$%
.The maximum excursion away from the unstable fixed point will be given by $%
C=0$%
\[
\zeta ^{2}=\frac{|\beta |^{2}}{\rho (n,Z)}. 
\]

We see then that the sharp resonances can make this excursion small, which
is consistent with a sharp frequency emited by the tangent dynamics.

Let us briefly discuss the shape of the level curves of such a constant:
Notice that we have involved the frequency of the $z$ coordinate oscillation
mode in the resonances of section VI, which makes the level curves of the
constant involve displacements along the z-direction. These level curves
will then have the shape of a torus encircling the circular orbit, and with
an excursion amplitude given by the above formula. When $\rho (n,Z)>0$, the
six constants intersect to confine a bounded manifold inside the resonance
islands. In quantum mechanics this information is provided by the
normalizability of the corresponding wave function, and the boundedness of
the region inside the resonance islands seems to be the analogous concept
here. We study the detailed stability problem in another publication\cite
{deluconvex}

I acknowledge discussions with R. Napolitano, L.Galgani , A. Carati, A.
Giorgilli and D. Bambusi . This work was supported by Fapesp.

\clearpage

\begin{tabular}{|c|c|c|}
\hline
$
\begin{array}{l}
\text{Coordinate} \\ 
\text{Involved}
\end{array}
$ & Coulombian roots & Corrected roots \\ \hline
$\delta z^{d}$ & $\lambda =\pm \frac{1}{2}$ & $\lambda =\pm \frac{1}{2}[1-(%
\frac{2-5\phi }{4})|\beta |^{2}]$ \\ \hline
$\left( \delta z^{\alpha },\delta z^{r}\right) $ & $\lambda =0,0,\pm \sqrt{%
2\Delta }$ & $\lambda =0,0,\pm \sqrt{2\Delta }\left( 1-\frac{3}{2}\phi
|\beta |^{2}\right) +\frac{64i}{3}Z\phi ^{2}|\beta |^{3}$ \\ \hline
$\delta {\bf x}^{d}$ & $\bar{n}=0,-1,-\frac{1}{2},-\frac{1}{2}$ & $\bar{n}=-%
\frac{7+6\phi }{16}|\beta |^{2},-1+\frac{7+6\phi }{16}|\beta |^{2},-\frac{1}{%
2}\pm \sqrt{\frac{3+6\phi }{2}}|\beta |$ \\ \hline
$\left( \delta {\bf x}^{\alpha },\delta {\bf x}^{r}\right) $ & $
\begin{array}{l}
\bar{n}=0,0,-1,-1, \\ 
\frac{-1}{2}\pm \sqrt{\frac{1}{4}-\Delta +\sqrt{\Delta (9\Delta -1)}}
\end{array}
$ & $
\begin{array}{c}
\bar{n}=0,0,-1,-1, \\ 
-\frac{1}{2}\pm \sqrt{\frac{1}{4}-\Delta +\sqrt{\Delta (9\Delta -1)}}\pm
c(Z)|\beta |^{2}+id(Z)|\beta |^{3}
\end{array}
$ \\ \hline
\end{tabular}

\begin{figure}[tbp]
\caption{Table 1: Corrections to the sixteen stable regular roots, $%
\phi\equiv 1/(8Z-2)$ and $\Delta\equiv Z\phi (1+2y)$.}
\end{figure}

\end{document}